\documentclass[12pt,preprint]{aastex}

\shorttitle{On the Weakening of Chromospheric Magnetic Field}
\shortauthors{Nagaraju, Sankarasubrmanian \& Rangarajan}

\begin{document}

\title{ON THE WEAKENING OF CHROMOSPHERIC MAGNETIC FIELD IN ACTIVE REGIONS} 

\author{K. Nagaraju}
\affil{Indian Institute of Astrophysics, Koramangala, Bangalore-560034, India; nagaraj@iiap.res.in}

\author{K. Sankarasubramanian}
\affil{ISRO Satellite Center, Airport road, Monera, Bangalore-560017, India;
sankark@isac.gov.in}

\and

\author{K. E. Rangarajan}
\affil{Indian Institute of Astrophysics, Koramangala, Bangalore-560034, India;
rangaraj@iiap.res.in}

\begin{abstract}
Simultaneous measurement of line-of-sight (LOS) magnetic and 
velocity fields at the photosphere and chromosphere are presented.
Fe I line at $\lambda6569$ and $H_{\alpha}$ at $\lambda6563$ 
are used respectively for
deriving the physical parameters at photospheric and chromospheric heights.
The LOS magnetic field obtained through the center-of-gravity 
method show a linear relation between photospheric and chromospheric field 
for field strengths less than $700$ G. 
But in strong field regions, the LOS magnetic field values derived 
from $H_{\alpha}$ are much weaker than what one gets from the linear relationship
and also from those expected from the extrapolation of the photospheric 
magnetic field.
We discuss in detail the properties of magnetic field observed in $H_{\alpha}$
from the point of view of observed velocity gradients.
The bisector analysis of $H_{\alpha}$ Stokes $I$ profiles show larger velocity 
gradients in those places where strong photospheric magnetic fields 
are observed. These observations may support the view that the stronger fields
diverge faster with height compared to weaker fields.

\end{abstract}

\keywords{Sun: magnetic fields - Sun: photosphere - Sun: chromosphere - sunspots}

\section{INTRODUCTION}

It is inferred through various observations that the magnetic field is 
playing a central role in the solar energetics that take place in the
higher layers of the atmosphere (see for eg. \citet{regnier06}).
But, it is somewhat difficult to  
obtain a reliable 
vector magnetic field measurements at the chromosphere and corona 
\citep{hector05, judge07}.
The measurements of the photospheric magnetic fields are 
relatively well established. 
However, simultaneous vector magnetic field observations at the 
photospheric and chromospheric heights will give a better handle on the 
understanding of the magnetic structuring of the solar atmosphere.
More direct and reliable measurements of the magnetic field at the chromosphere
will also serve as a better boundary condition for extrapolating the magnetic 
fields to the coronal heights.

Measurements of the vector magnetic field in $H_{\alpha}$ is particularly
important in understanding the connection between photospheric, chromospheric
and coronal magnetic field as its height of formation ranges almost
from the upper photosphere to upper chromosphere \citep{vernaza81}.
Also, this is one of the most widely used spectral lines for the
study of solar chromosphere (see a recent review by \citet{rutten07}).

Comparison of the active region magnetic fields measured in $H_{\alpha}$
and in the photosphere show a one-to-one correspondence in the weak field regions.
But in the strong field regions, like umbra, there is a 
considerable deviation from the linearity \citep{bala04, hanaoka05}. 
Infact the LOS magnetic field measured in $H_{\alpha}$ weakens much faster
for the corresponding strong field regions of the photosphere. 
At this point it may not be appropriate to demarcate the field strength 
above which this deviation happens because different observations show 
different deviation points. \citet{gosain03} have also reported the systematic 
weakening of the magnetic 
field derived from Mg I $\lambda5173/5184$ lines in comparison with that of
Fe I $\lambda6301.5/6302.5$ lines. In their case, the magnetic field
measured by Mg I lines agree with the potential field extrapolation of the photospheric LOS magnetic field in
weak field regime. In strong field regions there is a systematic shift 
towards lower values but still linear.
Simultaneous observations of He I at $\lambda10830$ and Si I lines at $\lambda10827.1$ by \citet{choudhary02} suggest that
the field diverges faster in the upper layers of the chromosphere.

However there is no conclusive explanation for the observed 
weaker chromospheric magnetic fields for the corresponding 
strong photospheric fields. 
Various possibilities have been discussed by \citet{bala04} \&  
\citet{hanaoka05} about weaker chromospheric fields observed in umbra.
\citet{hanaoka05} have discussed the possibility of scattered light and/or peculiarity of the 
atmosphere (radiative transfer effects) for the decrease in 
the polarization signal which in turn cause the underestimation of the 
magnetic field. 
\citet{bala04} have suggested that the strongest fields measured at the photosphere
diverge spatially and more quickly than the weak fields when propagating upward.
In this paper, we address these issues from the point of view of the observed 
velocity gradients.

It is well known that, velocities and magnetic fields are coupled to each other 
in the solar atmosphere (see for eg. \citet{rajaguru06}).
Any change in the magnetic field is expected to alter the plasma motion and hence 
the observed velocity. 
If the field strength decreases with height then the velocity is expected to
increase and vice versa, for the simple reason that the inhibition of the
plasma depends on the magnetic field strength \citep{spruit79}.
It is also well known that the magnetic field strength in the umbral region at the 
photosphere is large. 
If the magnetic field measured with $H_{\alpha}$ in the 
umbral region is weaker, then it shows that the magnetic field gradient is 
larger than expected which  implies larger velocity gradients in the 
umbral region compared to the penumbral region at chromospheric heights.
In order to observationally verify this we have analyzed the velocity and magnetic 
fields estimated from the simultaneous spectropolarimetric
observations of an active region using $H_{\alpha}$ and Fe I spectral lines.

The spectropolarimetric observations which were carried out at the Kodaikanal 
solar telescope using a dual beam polarimeter are discussed 
in section(\ref{sect:instr&obs}). 
Sections (3) and (4) discuss the data reduction and analysis 
procedures respectively. 
A comparison of LOS magnetic field at the chromosphere and photosphere is presented
in section (\ref{sect:losmag}). Velocity gradients obtained through bisector technique 
are discussed in section (\ref{sect:velgrad}).

\section{INSTRUMENT AND OBSERVATIONS}
\label{sect:instr&obs}

Spectropolarimetric observations were carried out using a newly added dual beam 
polarimeter to the spectrograph at Kodaikanal solar telescope (\citet{nagaraj07} and 
also see \citet{bappu67} for details about the spectrograph and telescope setup).  
The wavelength region of the observations presented in this paper
includes $H_{\alpha}(\lambda6563)$ and Fe I ($\lambda6569$) lines. 
Both $H_{\alpha}$ and Fe I lines are magnetically sensitive
with effective Land$\acute{e}$ 'g' factor of 1.048 \citep{casini94}
and 1.4 \citep{kobanov03} respectively.
The measured instrumental broadening and the linear dispersion close to this 
wavelength region in second order of diffraction are $38\pm0.5 m\AA$ and 
$10.15m\AA/pixel$ respectively.
The wavelength calibration was done using the telluric lines ($H_2O$)  at
$\lambda6570.63$ and $\lambda6561.097$ \citep{moore66}.

An eight stage modulation scheme was used for the measurement of the general 
state of polarization \citep{nagaraj07}.
In this modulation scheme the measurement of the Stokes parameters are well 
balanced over the duration of eight stages of intensity measurements.
The rotation of the retarders were done manually for the modulation of
the input light.

The spectropolarimetric data of an active region NOAA0875 presented in 
this paper were obtained on 28th April, 2006.
The heliographic coordinates of the sunspot during observations were
$11^o$ south and $18^o$ west.
Scanning of the sunspot was done by moving the Sun's image
in an east-west direction in steps of $\approx 5''$.
For each slit position, the modulated intensities were recorded by
the CCD detector. The CCD is a $1K \times 1K $ Photometrics detector with 
the pixel size of $24 \mu$. 
Eight stages of modulation took about $90$ s with a typical exposure time 
of $0.5$ s.

\section{DATA REDUCTION}

The spectral images were corrected for dark current and gain table 
variation of the pixels (details about the flat fielding of spectropolarimetric data
can be found in \citet{schl02,beck05}). 

Model independent velocity and magnetic field gradients can be derived
from the bisector technique for the range of heights over which the 
spectral line is formed, provided the full spectral profile is available without any
blend \citep{bala97, sankar02}.
The application of bisector technique to $H_{\alpha}$ is restricted
because of the blend in its red wing by the Zeeman sensitive 
Co I ($\lambda6563.4$) line which forms at the photospheric heights. 
In the following section we discuss the procedure to remove this blend
and its limitations.

\subsection{Blend Removal Procedure}
\label{sect:blendrem}

\begin{figure}[h]
\plotone{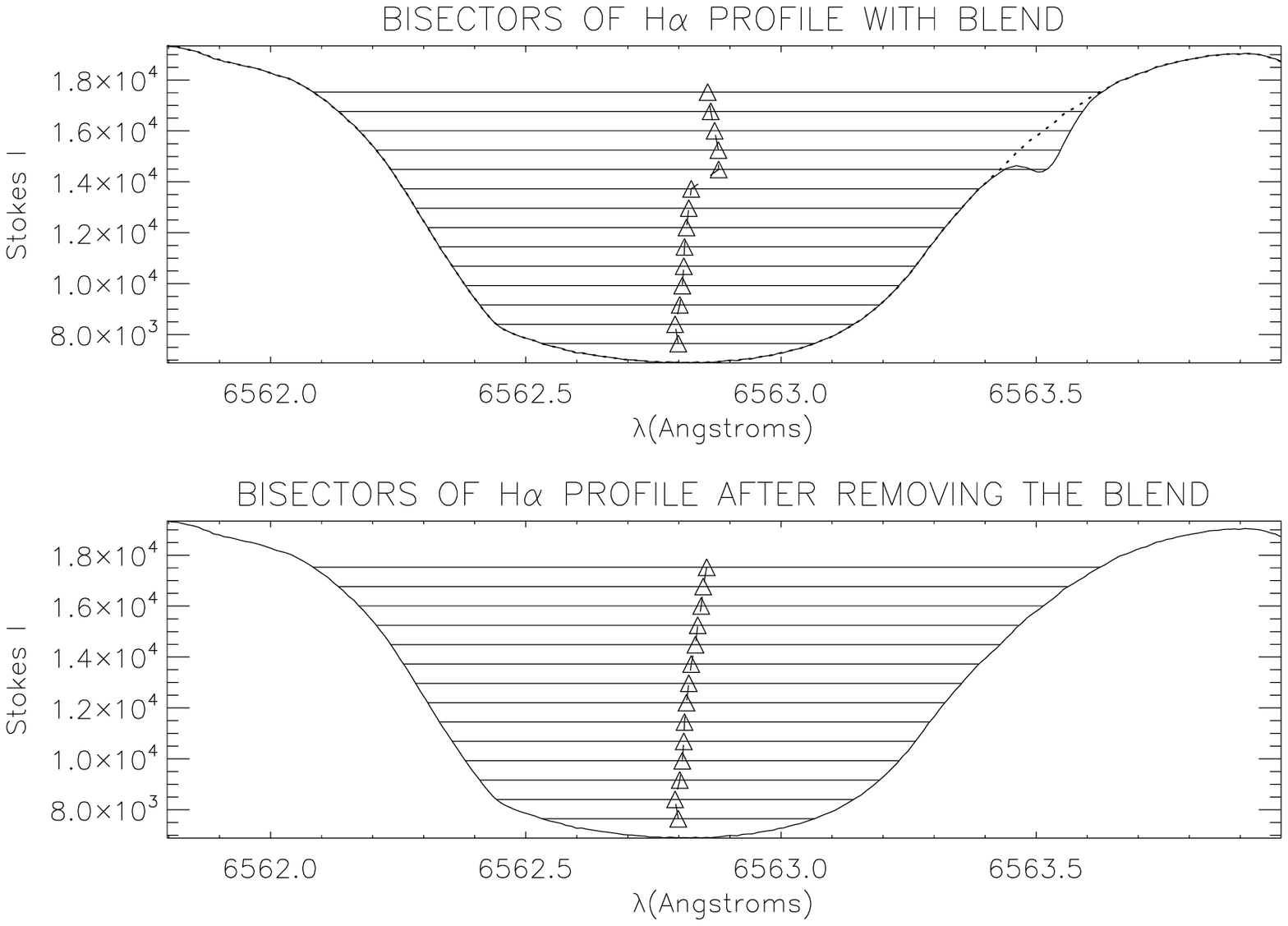}
\caption{Typical $H_{\alpha}$ Stokes $I$ profile with the bisectors
marked in diamond symbols. Dotted line in the top panel shows
the spectral region of the profile after removing the blend.}
\label{fig:blendrem}
\end{figure}

Individual profiles of $H_{\alpha}$ corresponding to the orthogonally
polarized beams in each stage of modulation were considered for removing the 
blend. A function which is a linear combination of a Gaussian and a quadratic term was
fitted to the blend region of the observed $H_{\alpha}$ profile.
This non-linear least square fit (available in IDL) takes into account the curvature in the 
intensity profile of $H_{\alpha}$ line along with the Co I line profile 
approximated as a Gaussian. The Gaussian, constructed out of the fitted parameters,  
was removed from the observed profile.
Then the eight intensity measurements were combined to obtain 
Stokes $I,~Q,~U$ and $V$ spectral images through the demodulation 
procedure explained in \citet{nagaraj07}.
A typical $H_{\alpha}$ intensity profile before and after the blend removal 
is shown in Fig. \ref{fig:blendrem} with the bisectors marked
as diamond symbols. 

Even though it appears in the total intensity profile that
the effect of Co I line is completely removed, the blend residuals
still appear in Stokes $Q,~U$, and $V$ profiles.
The residuals are due to the Gaussian approximation used for the Co~I line profile. 
However, it is demonstrated in the section \ref{sect:velgrad} that the blend 
residuals do not have any effect on the velocity gradients calculated 
from the $H_{\alpha}$ intensity profiles.

The better way to remove the blend may be to synthesise the Stokes profiles of Co I using radiative transfer equations
along with the atmospheric parameters obtained through Fe I ($\lambda6569$) line.
However in this paper only Stokes $I$ profiles
are considered for the velocity gradients estimation and a restricted
spectral range of $H_{\alpha}$ about line center for the
LOS magnetic field estimation. Hence, the simple blend removal method outlined in this paper
is found to be sufficient.

\subsection{Correction for Polarimeter Response and Telescope Induced
Cross-talks}

The polarimetric data was corrected for polarimeter response
(for details see \citet{nagaraj07}).
The instrumental polarization introduced by the telescope was corrected
by using the telescope model developed by \citet{bala85} and \citet{sankar00}.
The refractive index values for Aluminium coating
used in the model are obtained from the catalog \citep{walter78}. 
Since these values may 
be different from the actual values, there still remain residual cross-talks
among Stokes parameters. These residual cross-talks are removed by the
statistical method given in \citet{schl02}.

\section{DATA ANALYSIS}
\subsection{LOS Magnetic Field}
\label{sect:losb}

The LOS magnetic fields at the photospheric and chromospheric
heights are derived using the Fe I line  and
$H_{\alpha}$ respectively. For the derivation of 
the LOS magnetic fields, center-of-gravity (COG) method was used 
\citep{rees79, cauzzi93, han03, bala04}.

For the COG method, the LOS field strength is given by

\begin{equation}
B_{LOS} = \frac{(\lambda_+ - \lambda_-)/2}
          {4.667\times 10^{-13} \lambda_0^2g_L},
\end{equation}

where $\lambda_0$ is the central wavelength of the line in $\AA$, $g_L$ is
the Land$\acute{e}$ `g' factor of the line, and $\lambda_{\pm}$ are the
COG wavelengths of the positive and negative circularly polarized
components respectively. The COG components are calculated as

\begin{equation}
\lambda_{\pm} = \frac{\int [I_{cont} - (I \pm V)] \lambda d\lambda}
                     {\int I_{cont} - (I \pm V)d\lambda}
\end{equation}

where $I_{cont}$ is the local continuum intensity. The integration is over
the spectral range of a given spectral line. Since Fe I line is free of blend,
the full spectral range is available for the COG method.
For the $H_{\alpha}$ line, the spectral range is restricted due to the blending 
of the Zeeman sensitive Co I line. Even though the blend is removed in Stokes $I$ 
profile using the Gaussian fit technique explained in the section (\ref{sect:blendrem}),
there still remains a residual in polarization profiles. To avoid introducing 
any artifacts in the LOS magnetic field values derived using COG method, a restricted 
spectral range about the $H_{\alpha}$ line core is used. This spectral region was 
selected by looking at the strongest Stokes $V$ signal of Co I line. Hence the derived
LOS magnetic field using $H_{\alpha}$ would correspond mostly to the upper 
chromosphere \citep{rutten07}. 

The expected maximum underestimation of the photopsheric LOS magnetic field 
is about $12\%$ for the kind of field strengths presented in this 
paper. 
Because of the large intrinsic Doppler width of $H_{\alpha}$, 
the underestimation of the
LOS field is not expected even for the strongest magnetic field
observed at the chromosphere \citep{han03}.
However we checked the reliability of the COG method in deriving LOS field
from $H_{\alpha}$ by applying it to synthetic Stokes  $I$ and $V$ profiles 
of $H_{\alpha}$ obtained using the radiative transfer code of \citet{han98}.
From these studies it was found that the COG method overestimates the LOS field
by about $1.5\%$ in the field strength range 0 to 2000 G.
However, the main error involved in the determination of the LOS magnetic field, 
for this observation, is due to
the error in estimating the shifts between the COG wavelength 
positions $\lambda_+$ and $\lambda_-$ which is expected to be 
less than 70 G.

\subsection{Photospheric Vector Magnetic Field}
\label{sect:vmag:photo}
The strength and the orientation of the chromospheric magnetic field are
very much dependent on its vector nature at the photosphere
(\citet{wiegelmann06} and references there in).
Inference of the vector magnetic field at the chromosphere from the 
$H_{\alpha}$ observations is difficult because, 
Stokes  $Q$ and $U$ are hardly discernible.
Only at few locations of penumbra they are above the noise level 
($2 \times 10^{-3} I_{cont}$ in our observations).
Hence for the correct interpretation of the LOS magnetic field obtained 
from $H_{\alpha}$ it is important to have the complete information 
about the vector magnetic field at the photosphere.

The photospheric vector magnetic fields are obtained by inverting the observed Stokes profiles of Fe I line. 
Milne-Eddington Line Analysis using a Numerical Inversion 
Engine (MELANIE) 
\footnote{http://www.hao.ucar.edu/public/research/cic/index.html}  
was used to perform the inversion.
 MELANIE performs 
non-linear least-square fitting of the observed Stokes profiles under 
Local-Thermodynamic-Equilibrium (LTE) condition by assuming Milne-Eddington atmosphere.
Inversion code returns magnetic field strength, inclination angle with
respect to LOS, azimuth, line strength, damping parameter,
LOS velocity, source function and its gradient with optical depth,
macroturbulence and fraction of stray light/fill factor of the non-magnetic
component. 

The fit error in estimating the magnetic field ranges from 
$50$ G for the Stokes profiles which are symmetric and well above
the noise level to $250$ G for the Stokes profiles which are highly 
asymmetric and close to noise level. The maximum fit error in the estimation of 
the field orientation is about $5^o$ and its azimuth is $6^o$.  
Other physical parameters returned by the MELANIE are not used in the 
current work and hence are not discussed here.

\subsection{Velocity Gradients}
The measurement of the magnetic field in $H_{\alpha}$ has been 
difficult to interpret \citep{bala04, hanaoka05}.
However it is possible to address some of these difficulties from the
study of plasma motions as magnetic field and plasma motion 
influence one another in the solar atmosphere \citep{gary01}.
Hence the study of velocity and its gradient may help in better
interpretation of the observed magnetic field.

The velocities at the photosphere and chromosphere are obtained through
COG method \citep{han03}.
The COG wavelength $\lambda_{COG}$ of a line profile $I$ is defined as the
centroid of its residual intensity profile:
\begin{equation}
\lambda_{COG} = \frac{\int \lambda(I_{cont}-I)d\lambda}
{\int (I_{cont}-I)d\lambda}.
\end{equation}
The LOS velocity with respect to the average quiet Sun 
reference ($\lambda_{ref}$) is defined as
\begin{equation}
v_{LOS} = \frac{c(\lambda_{COG}-\lambda_{ref})}{\lambda_{ref}},
\end{equation}
where $c$ is the speed of the light.

Bisector technique has been applied to Stokes $I$ profiles 
to derive the velocity gradients both at
the photosphere and chromosphere.
The Stokes $I$ profiles of Fe I and $H_{\alpha}$ are separately considered
for bisector analysis.
Bisectors are obtained at 9 equal intensity levels between line core and the wing for 
Fe I and 14 equal intensity levels between line core and wing for $H_{\alpha}$ respectively.
Out of 9 bisectors of Fe I line only 7 are considered, namely, the bisectors between second
and eighth counting from the line core. Similarly for $H_{\alpha}$ the bisectors between 
second and thirteenth are considered totalling 12 bisectors. The bisectors very close to line
core and wing are not considered because of the lower signal-to-noise ratio and to avoid the 
influence of the continuum respectively. For the Fe I line, the wavelength position of the 
second bisector which corresponds to higher atmospheric layer was subtracted from the seventh
bisector which corresponds to lower atmospheric layer. Similarly for $H_{\alpha}$, the wavelength
position of the second bisector was subtracted from the thirteenth bisector. 
The wavelength differences
($\Delta\lambda$s) thus obtained are converted into velocities - which would then represent the 
velocity difference between the lower and higher atmospheric layers - using the
following relation,
\begin{equation}
\label{eq:veldifbisec}
\Delta V_{bs} = \frac{c \Delta\lambda}{\lambda_0}.
\end{equation}
Where $\lambda_0$ is the rest wavelength of the spectral line under 
consideration. The velocity difference defined
in Eq. (\ref{eq:veldifbisec}) represents the velocity gradient over the line 
formation height.  

The errors in estimating the velocity differences are mainly due to the
errors involved in finding the wavelength shifts. 
The maximum error in estimating
the velocity gradients is about $0.09$ km s$^{-1}$.

\subsection{Stokes  $V$ Amplitude Asymmetry}
Amplitude and area asymmetries of Stokes  $V$ profiles are caused
by the gradients in the velocity and magnetic fields 
(see for eg. \citet{sanchez92, sankar02}). Hence their analysis will give
us some handle on the understanding of the field gradients.

If $a_r$ and $a_b$ represent the amplitudes of red and blue wings of Stokes $V$ 
respectively then the amplitude asymmetry is defined as,

\begin{equation}
\delta a = \frac{|a_b|-|a_r|}{|a_b|+|a_r|}.
\end{equation}

The area asymmetry is not considered in this paper due to the difficulty in estimating
it for the $H_{\alpha}$ in the presence of Co I line blend.

\section{RESULTS AND DISCUSSIONS}

\subsection{Comparison of Photospheric and Chromospheric LOS Magnetic Fields}
\label{sect:losmag}

The scatter plot of LOS magnetic field derived from $H_{\alpha}$ and Fe I is 
shown in Fig. \ref{fig:28fld}. The plot shows that the chromospheric
magnetic field is weaker in general compared to its photospheric counterpart. 
The important point to note from this figure is that the chromospheric
fields are much weaker in the locations where the strong photospheric fields 
are observed. 
Similar kind of observations were reported earlier by \citet{bala04} using 
simultaneous observations in $H_{\alpha}$  and Fe I line at $\lambda6301.5$ and 
\citet{hanaoka05} who compared the LOS field measured in $H_{\alpha}$ with the 
magnetograms of SOHO/MDI.
These observations may imply that the stronger fields weaken much faster
when they propagate upward in the solar atmosphere. 

To illustrate the quicker weakening of the stronger fields, plots of the 
photospheric and the chromospheric LOS field strengths along two different 
radial slices of the sunspot are shown on the right side of the Fig. \ref{fig3}.
The radial slices considered for this purpose are marked as 1 and 2 on
the SOHO/MDI intensitygram showing the sunspot analysed in this paper 
\footnote{Because of the coarser step size used for scanning the sunspot
($\approx$ 5 arc sec), the raster images do not give a good representation of
the observed region and also there was no imaging facility available during
these observations and therefore intensitygram from SOHO/MDI is used. 
The radial cuts marked on the image represent the approximate slit
positions of the observations.}.
The top panel on the right side of the Fig. \ref{fig3} shows the plots
of photospheric and chomospheric LOS magnetic field strengths along the
radial cut (marked as 1 on the sunspot image) passing close to the central 
umbra. Note that, along this radial cut the photospheric field strength
increases systematically from the penumbral region to the umbral region. 
While the chromospheric field strength increases upto about 800 G, more or 
less linear with the photospheric field, and then starts decreasing towards the 
umbra. The decrease in the chromospheric field is much larger close to central 
umbra.  The photospheric field along the radial cut 2 as shown in the bottom
panel on the right side of the Fig. \ref{fig3} shows the behavior very
similar to that was seen along the radial cut 1. That means the photospheric
field strength is larger in the umbral region and decreases towards
the edges of the sunspot. In the case of chromospheric field strength,
the values do not decrease towards the umbra as was seen for the radial cut 1
but, they are considerably smaller compared to its photospheric counter part.
The main difference between these two radial cuts is that the field strength
in the umbral photosphere for radial cut 1 is larger compared to that of 
radial cut 2.
This may be an indication of the faster divergence of the stronger fields.
However, there are other possibilities which can cause this observed weaker fields
in the umbral chromosphere and they are : weakening of the polarization signal due to 
scattered light within the instrument and from nearby quiet Sun; peculiarity of 
the atmosphere such as discussed by \citet{hanaoka05};
the methods used to derive LOS magnetic field (inversion of $H_{\alpha}$ 
Stokes profiles is yet to be established); 
inclination of the magnetic field; 
and most importantly the height of formation which
is highly ambiguous \citep{hector04}.
We will discuss these issues in detail after looking at the results from the 
analysis of velocity gradients both at the photosphere and chromosphere.

\subsection{Velocity Gradients and the Nature of the Magnetic Fields}
\label{sect:velgrad}

The velocities calculated using COG method show a
typical behavior of Evershed flows both at the photosphere and chromosphere. 
That means the limb side penumbra shows red shift with 
respect to quiet Sun where as center side penumbra shows blue shift
at the photospheric heights. The situation is exactly opposite
at chromospheric heights which are consistent with the well known 
inverse Evershed effect.
COG velocities in umbral regions both at photosphere and chromosphere 
are smaller compared to penumbral regions.

The plots of bisector velocity differences (Eq. \ref{eq:veldifbisec})
as a function of photospheric magnetic field strength are shown
in Fig. \ref{fig:velgradfe} for the Fe I line.
Top panel in this figure is for all the points over
the total field of view (FOV), where as the bottom left and right panel is for umbral and
penumbral regions of the observed spot, respectively. 
Note from these figures that the large number of points correspond to
umbra have smaller velocity gradients than the penumbra.
Closer investigation of bisector velocity differences of Fe I line
show the flow pattern consistent with the well known Evershed flow.
That means larger portion of the limb side penumbra shows net downflow
(both core and wing side bisectors show redshifts)
and disk center side penumbra shows net upflow (both
core and wing bisectors show blue shifts with respect to the reference).
For the definition of the net up- and down-flow see \citet{bala97}.

We also found that the bisector velocity gradients and COG velocities 
observed in Fe I show a good correlation in agreement with the earlier 
observations.  That means, larger the COG velocity larger the velocity 
gradient at the photosphere.

In Fig. \ref{fig:vgradhalpha_full} the plots of bisector velocity 
gradients for $H_{\alpha}$  v/s the photospheric magnetic fields are shown.
Top panel in this figure is for the total FOV and left bottom panel for
umbral region and right bottom panel for penumbral region. 
Notice the increase in bisector velocity gradients measured in $H_{\alpha}$ 
with increase in the photospheric magnetic field strength.
Plots of velocity gradients along two radial slices of sunspot (marked as
1 and 2 in Fig. \ref{fig3}) also show that the larger velocity gradients at the 
chromosphere are located at which strong photospheric fields are observed 
(Fig. \ref{fig6}).  Top panels in Fig. \ref{fig6} show the plots of 
photospheric field strengths along the radial cut 1 and the radial cut 2. 
The corresponding plots of velocity gradients are shown in the bottom panels. 
Except at few locations (mostly in the limb side penumbra such as shown in the 
bottom right panel of Fig. \ref{fig6}) which show less variation in the  values 
of velocity gradients inspite of variation in the photospheric field 
strengths (which is evident also in Fig. \ref{fig:vgradhalpha_full}), most of 
the places the velocity gradients are larger where the photospheric magnetic 
field strengths are larger. 

Closer examination of the bisector wavelength positions 
of $H_{\alpha}$ Stokes $I$ profiles in different
regions of the sunspot with respect to the reference wavelength
reveals the following results (see figure 
\ref{fig:comprbisec}).

\begin{itemize}
\item{} In the limb side penumbra both line core and wing side bisectors 
show blue shift with respect to quiet Sun. The shifts in the line core 
bisectors are large compared to the line wing bisectors 
indicating the net upflow. 
\item{} In the umbral region also both line core side and wing side bisectors
show blue shifts with respect to the reference wavelength. The shifts in the 
line core side bisectors for the umbra are almost comparable to that of the 
limb side penumbra. However, the shifts in the line wing side bisectors for the umbra
are much smaller with respect to that of the limb side penumbra.
This would indicate again the net upflows but with larger velocity gradients.
\item{} In the center side penumbra the line wing bisectors show redshift in most of 
the places with respect to the quiet Sun. At few locations they show a small blue 
shift or no shift. The line core side bisectors show slight blue shift with respect
to quiet Sun reference wavelength position.
\end{itemize}
 
It is found through these analyses that the velocity gradients are larger 
in the umbra at the chromospheric heights compared to the penumbra 
(see Fig. \ref{fig:vgradhalpha_full}). 
Which is exactly in contrast with the flow pattern observed at the photosphere. 
The analysis of bisector velocity differences also indicate accelerated
upflows in the umbral region.

There is a little concern due to the residuals present after the Co I line
blend is removed. To make sure that there are no artifacts introduced due to
the residuals, the bisector velocity differences calculated by considering 
the spectral region of the line which is not affected by the blend are shown 
in Fig. \ref{fig:vgradhalpha_part} as a function of photospheric field 
strength. This figure also shows the trend that the velocity gradients 
increase with increase in photospheric magnetic field strength confirming 
the observations made with the blend removed full intensity profiles. As 
expected, the gradients are smaller due to the smaller wavelength regions 
considered in this case.  
	
To summarise, wavelength shifts analysis of Stokes $I$ bisectors show 
that the velocity gradients are larger in the umbral region than in
the penumbral region at the chromospheric heights.
Most importantly, accelerated upflows are observed in the umbral region.
In other words, LOS velocity increases upward more rapidly in
the umbral region than in the penumbral region at the chromosphere. 

Let us now address some of the possibilities, mentioned in the
beginning of this section, which can cause
the observed weaker chromospheric field in the umbral region from
the point of view of the velocity gradients. 

We found that the stray light
within the instrument is less than $2\%$ by comparing the quiet Sun
spectrum with the atlas \citep{wallace00}. This amount of stray light
is too small to account for the observed weaker Stokes $V$ signals
in the umbral region. Also, it can not account for the observed 
velocity gradients in $H_{\alpha}$.

\begin{figure}
\plotone{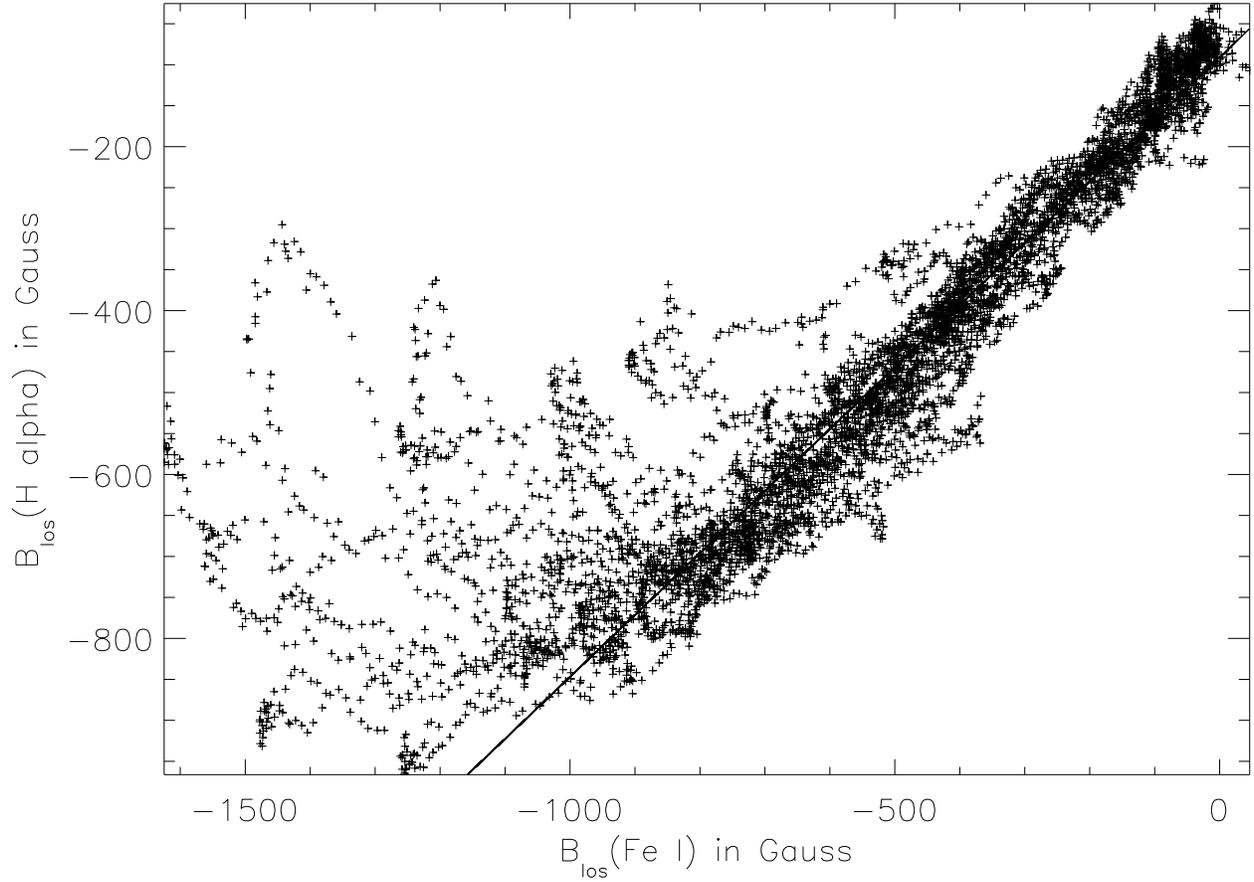}
\caption{Scatter plot of photospheric and chromospheric LOS magnetic field
derived using COG method applied to Fe I($\lambda6569$) and 
$H_{\alpha}$ respectively. Solid line is a linear fit to the field
strengths correspond to the penumbral region.}
\label{fig:28fld}
\end{figure}

The next question raised was about the reliability of the methods used to
estimate the LOS magnetic field. Despite the availability of more accurate
estimation of photospheric magnetic field through inversion, we
have used the COG method to maintain the uniformity while comparing
the LOS field at the photosphere and chromosphere (Fig. \ref{fig:28fld}).
This is because the inversion of $H_{\alpha}$ profiles to derive magnetic field
is yet to be established.
As discussed in section \ref{sect:losb} the underestimation of the LOS
field obtained from Fe I is expected to be more while the values
obtained from $H_{\alpha}$ will be less as confirmed through numerical simulations.
Also, the observed umbral fields are larger at the photosphere where as
they are smaller at the chromosphere. 
Hence the departure of LOS magnetic fields from the actual values
due to the method used to estimate them can not explain the
observed weakening of the magnetic fields. 

One of the possibilities which can cause the reduction in polarization 
(in effect the weaker magnetic field)
proposed by \citet{hanaoka05} is the peculiarity of the atmosphere.
However, the kind of peculiarity discussed by him can not explain the 
velocity gradients and the Stokes profiles of $H_{\alpha}$ discussed in this
paper.

It is a general wisdom gained from the extrapolation technique that the
magnetic topology at the chromosphere is very much dependent on the field
configuration at the photosphere. Extrapolation techniques like potential field
approximation suggest that larger the field inclination at the photosphere
chromospheic fields should also show larger inclination. 
From the inversion of Fe I line it was
found that the fields are oriented more close to LOS in the umbral region than
in the penumbral region. Hence we expect larger field orientation in 
penumbral region at the chromospheric heights also. 
As mentioned in section \ref{sect:vmag:photo}
$H_{\alpha}$ shows Stokes $Q$ and $U$ signals which are above the noise level
in penumbral region but not in umbral region indicating that the orientation
of the fields are larger in the penumbral region compared to the umbral region
even at the chromosphere. 
Hence we believe that the weaker LOS field observed in
the umbral region may not be due to the larger orientation angle with respect
to the LOS. This scenario is also verified using the velocity gradients observed
at the chromosphere, as larger inclination means smaller LOS velocity in
contradiction to the observed velocities (Fig. \ref{fig:vgradhalpha_full}).  

Another major difficulty in interpreting the $H_{\alpha}$ observations 
is the ambiguity in its height of formation. Studies based on the
response functions by \citet{hector04} show that the $H_{\alpha}$
is sensitive mostly to chromospheric magnetic field in the umbral model.
While in the quiet Sun model it shows sensitivity to both photospheric and
chromospheric magnetic fields. Since the magnetic fields at the photosphere
is large, the magnetic field measured in quiet Sun regions will mostly 
be photopsheric.
If the $H_{\alpha}$ were to show the sensitivity to magnetic field
in the penumbral model similar to that of quiet Sun model then one would expect
larger field values in the penumbra which will be an average of photospheric and
chromospheric fields while the umbral fields are exclusively chromospheric.
To confirm this, more forward modeling is needed which includes 
the comparison of $H_{\alpha}$ line formation at different regions
with different field configurations.
More over, response of $H_{\alpha}$ line to various physical
parameters needs to be studied and consistently explain
the observations such as velocity and velocity gradients presented
in this paper.

Other possibility which can cause the observed weaker chromospheric fields
in the umbral region is the faster divergence of stronger fields as suggested
by \citet{bala04}.  This scenario may consistently explain both the observed 
magnetic properties (Fig. \ref{fig:28fld} ) as well as the velocity properties 
(Figs. \ref{fig:vgradhalpha_full}/\ref{fig:vgradhalpha_part}).
Because of the decrease in the field strength, the plasma becomes less 
inhibited by the magnetic fields and hence more free to move.  That means if the
stronger fields diverge faster compared to weaker fields then the the velocities
should increase faster in the stronger field regions which are consistent with 
the observations.

\subsection{Stokes $V$ Amplitude Asymmetry}

From the close examination of Stokes $V$ profiles of $H_{\alpha}$ we found that the Co I 
line has not intruded up to the extent that the amplitudes are affected. 
Hence the results presented from $H_{\alpha}$ based on amplitude
asymmetry are reliable.

From the Fig. \ref{fig:ampasym_fe} it is clear that the amplitude asymmetry 
of Fe I line tends toward zero
with increase in photospheric magnetic field strength. This means, at the 
photospheric heights the gradients are smaller in umbra which corresponds 
to the region of strong fields. 
In contrast amplitude asymmetry observed in $H_{\alpha}$ tend to increase
with photospheric field strength as shown in the Fig. \ref{fig:ampasym_hi}.
This implies that the field gradients are larger in the umbra at the 
chromosphere compared to penumbra. 
Hence the analysis of amplitude asymmetry confirm the results obtained from the 
bisector analsysis in section \ref{sect:velgrad}.
	
\section{CONCLUSIONS}
\label{sect:concl}

In this paper we have used the multiwavelength spectropolarimetric tool to
understand the stratification of the magnetic and velocity fields in the solar
atmosphere. Out of the two lines considered for spectropolarimetry, 
one forms at photospheric height (Fe I $\lambda6569$) and the other spans
almost from the upper photosphere to upper chromosphere ($H_\alpha$). Hence, 
these lines were useful in studying the connection between physical parameters
at the photospheric and chromospheric heights. The main physical parameters
studied in this paper are the magnetic and velocity fields in an active region.

\begin{figure}
\plottwo{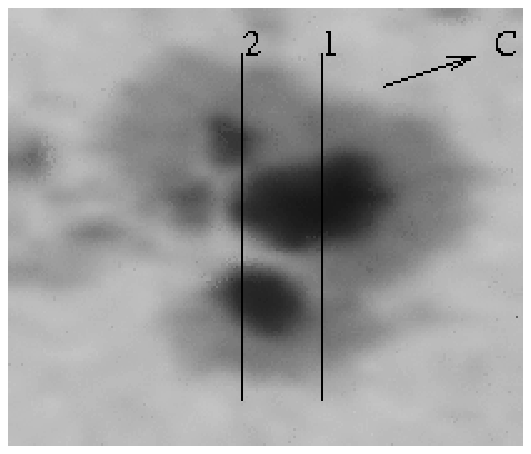}{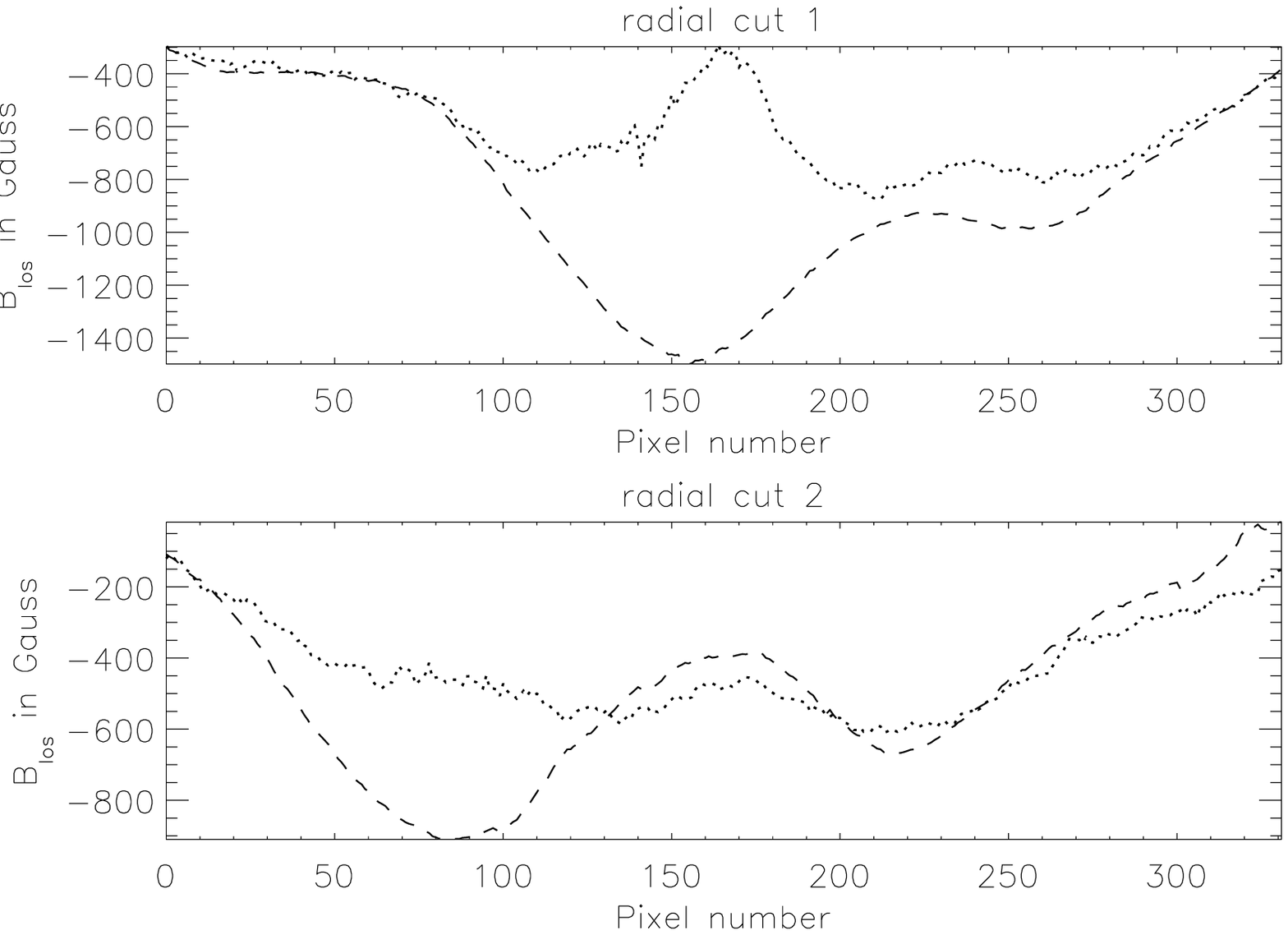}
\caption{On the right side of this figure shown are the plots of  LOS 
magnetic field strengths at the photosphere (dashed lines) and
the chromosphere (dotted lines) along two radial slices of the sunspot.
For reference, these two radial cuts are marked on the SOHO/MDI intensitygram. 
The arrow on the sunspot image indicates the disk center direction.}
\label{fig3}
\end{figure}

\begin{figure}
\plotone{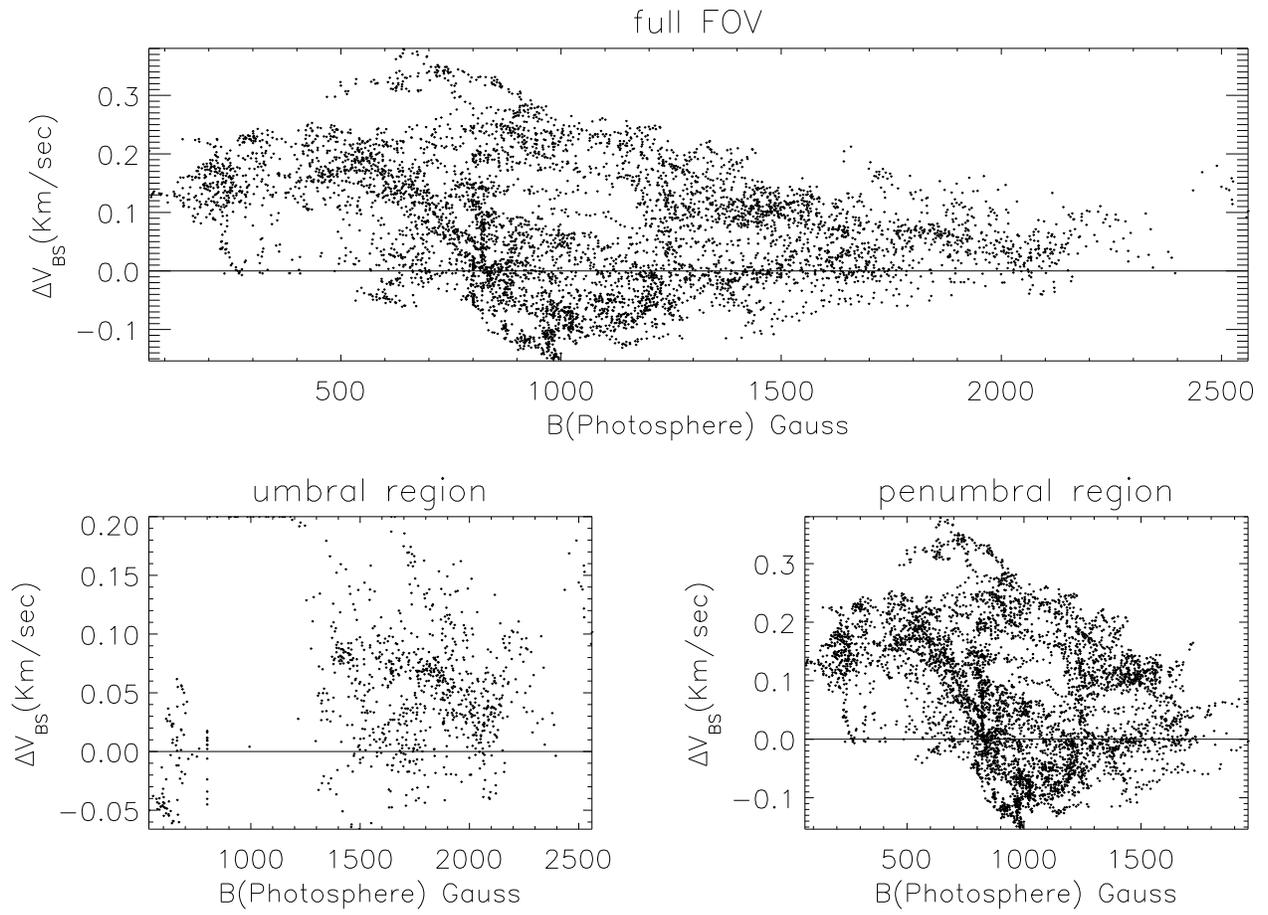}
\caption{Plots of velocity gradients at the photosphere 
 v/s photospheric magnetic field strength.}
\label{fig:velgradfe}
\end{figure}

\begin{figure}
\plotone{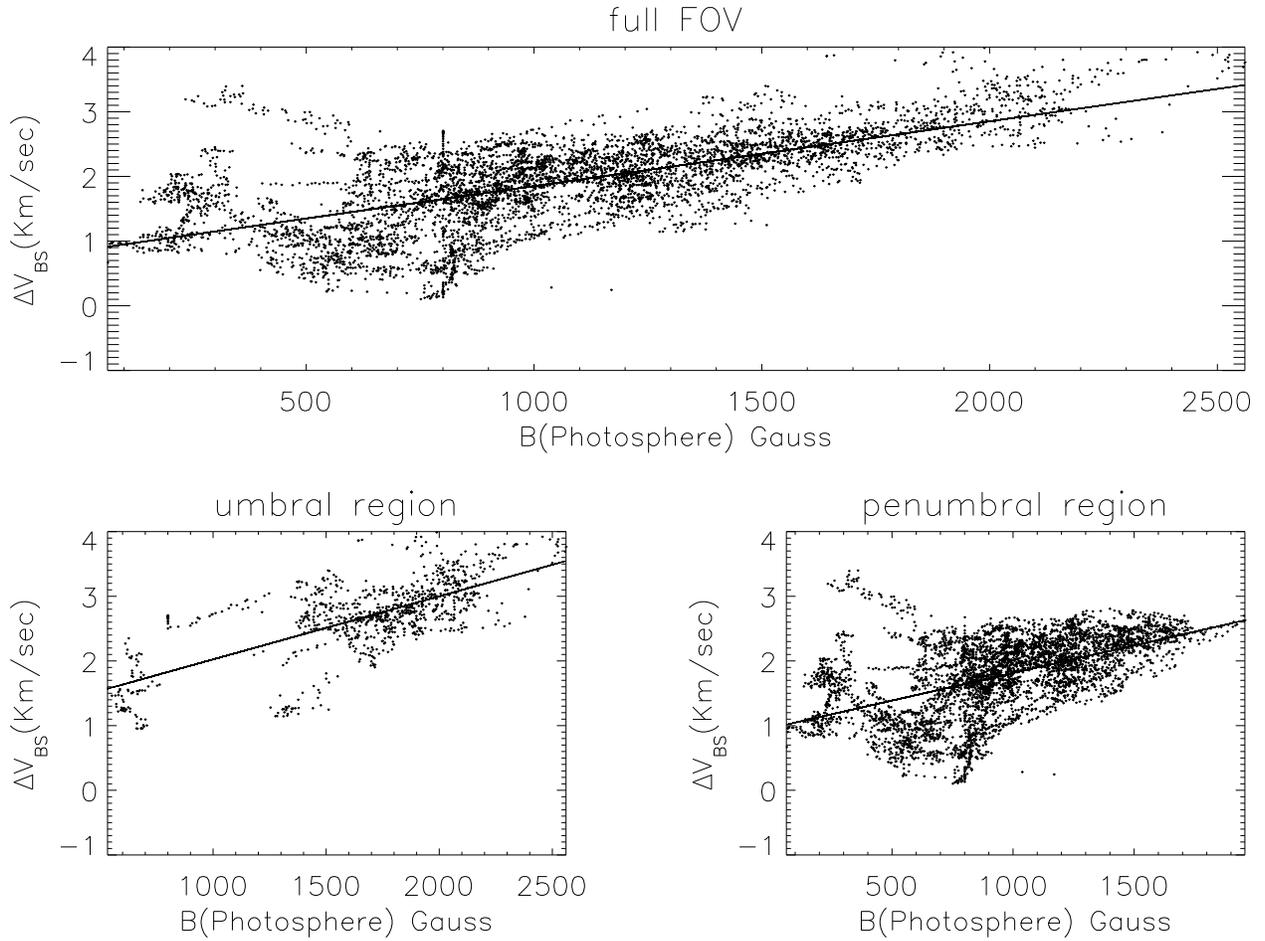}
\caption{Plots of velocity gradients at chromosphere derived using $H_{\alpha}$ with the bisectors considered for the full line profile
 v/s the total field strength (photospheric). Solid lines are the linear
fit to the data points.}
\label{fig:vgradhalpha_full}
\end{figure}

\begin{figure}
\plotone{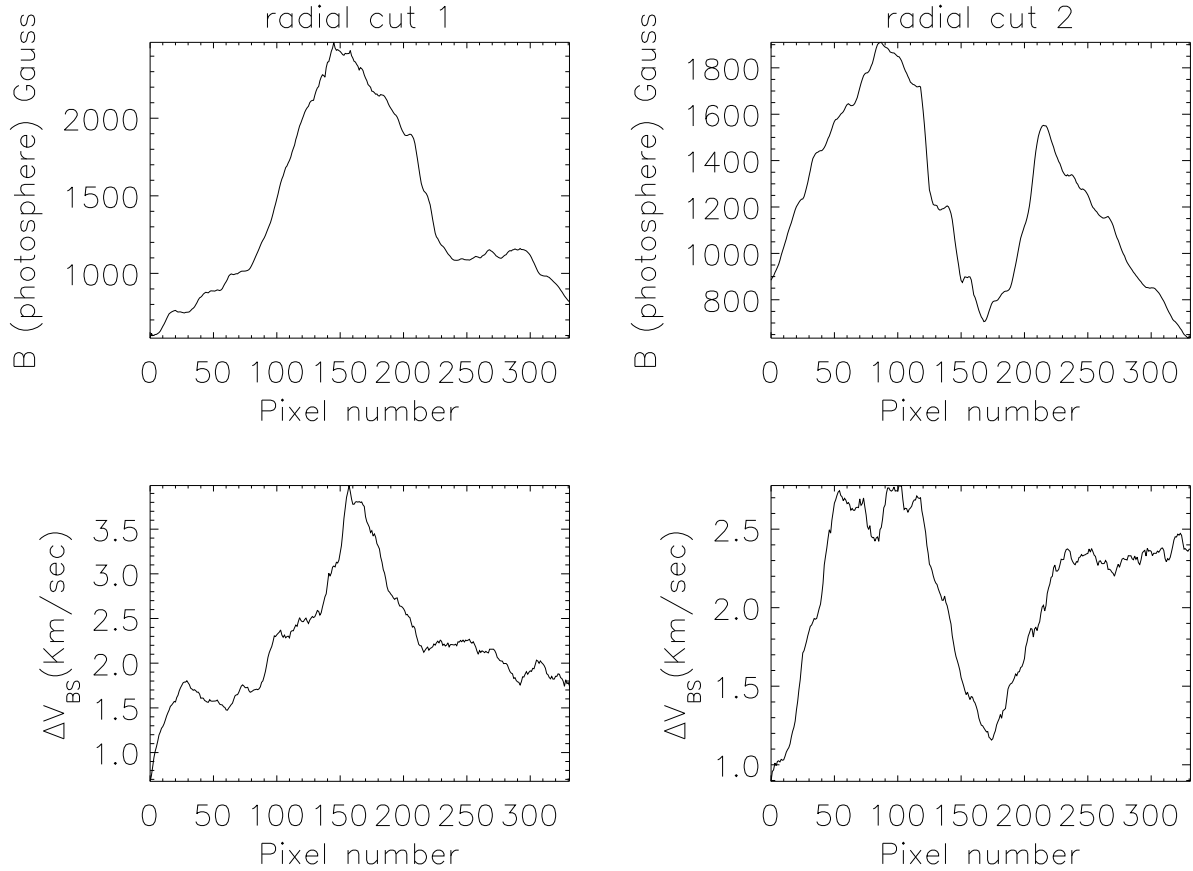}
\caption{The top panels in this figure show the plots of magnetic field
strength at the photosphere along two radial slices of the sunspot.
The radial slices are same as those shown in Fig. \ref{fig3}.
The corresponding plots of velocity gradients
at the chromosphere are shown in the bottom panels.}
\label{fig6}
\end{figure}

\begin{figure}
\includegraphics{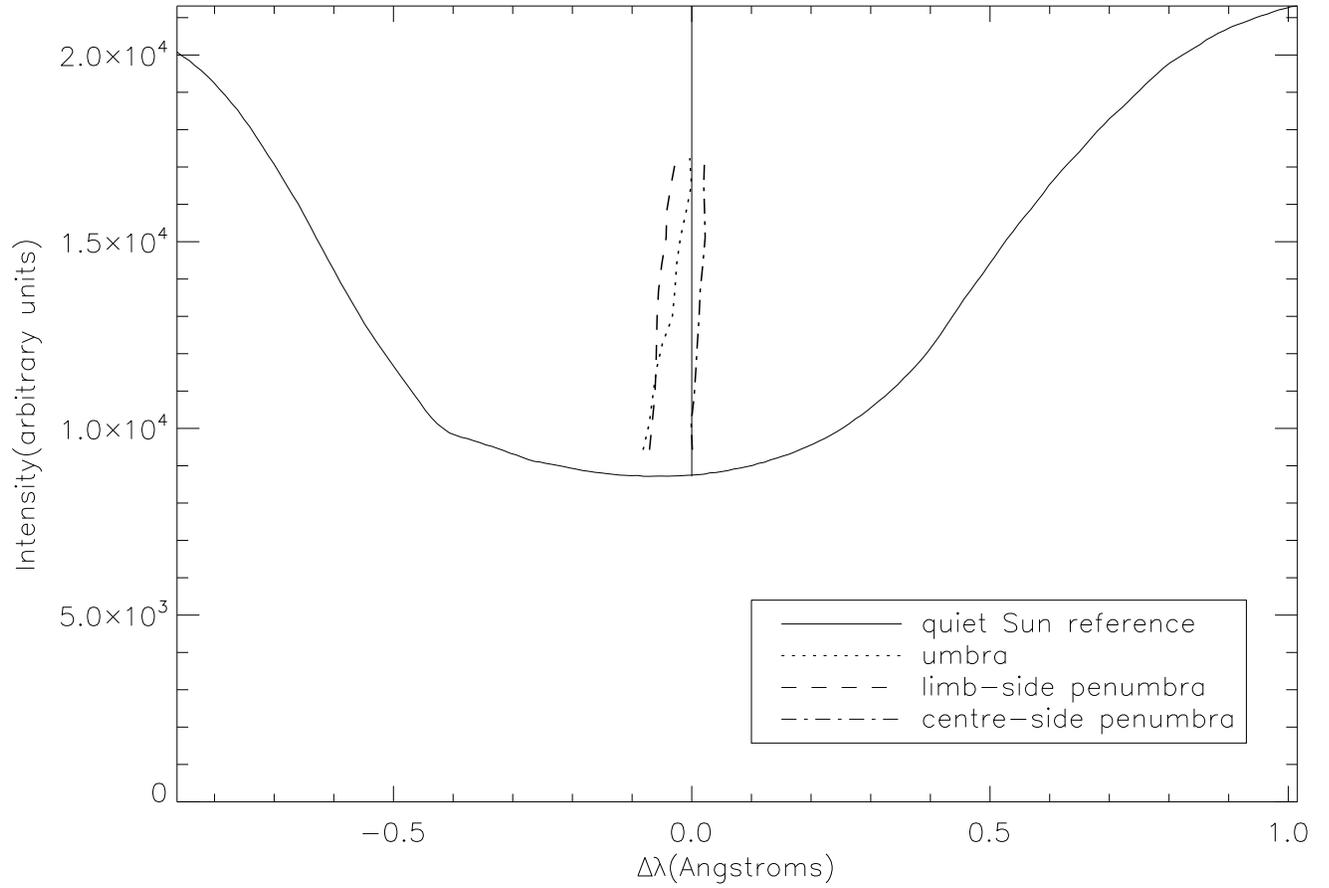}
\caption{Representative bisectors of $H_{\alpha}$ line profiles 
correspond to different regions of the sunspot are plotted on an 
average quiet Sun profile. 
The solid line represents the reference wavelength (which is a COG wavelength 
position of the nearby average quiet Sun profile). Dotted line represents
the bisectors location  typical of umbral profiles, dashed line-limb
side penumbra, dash dotted line-center side penumbra.}
\label{fig:comprbisec}
\end{figure}

\begin{figure}
\plotone{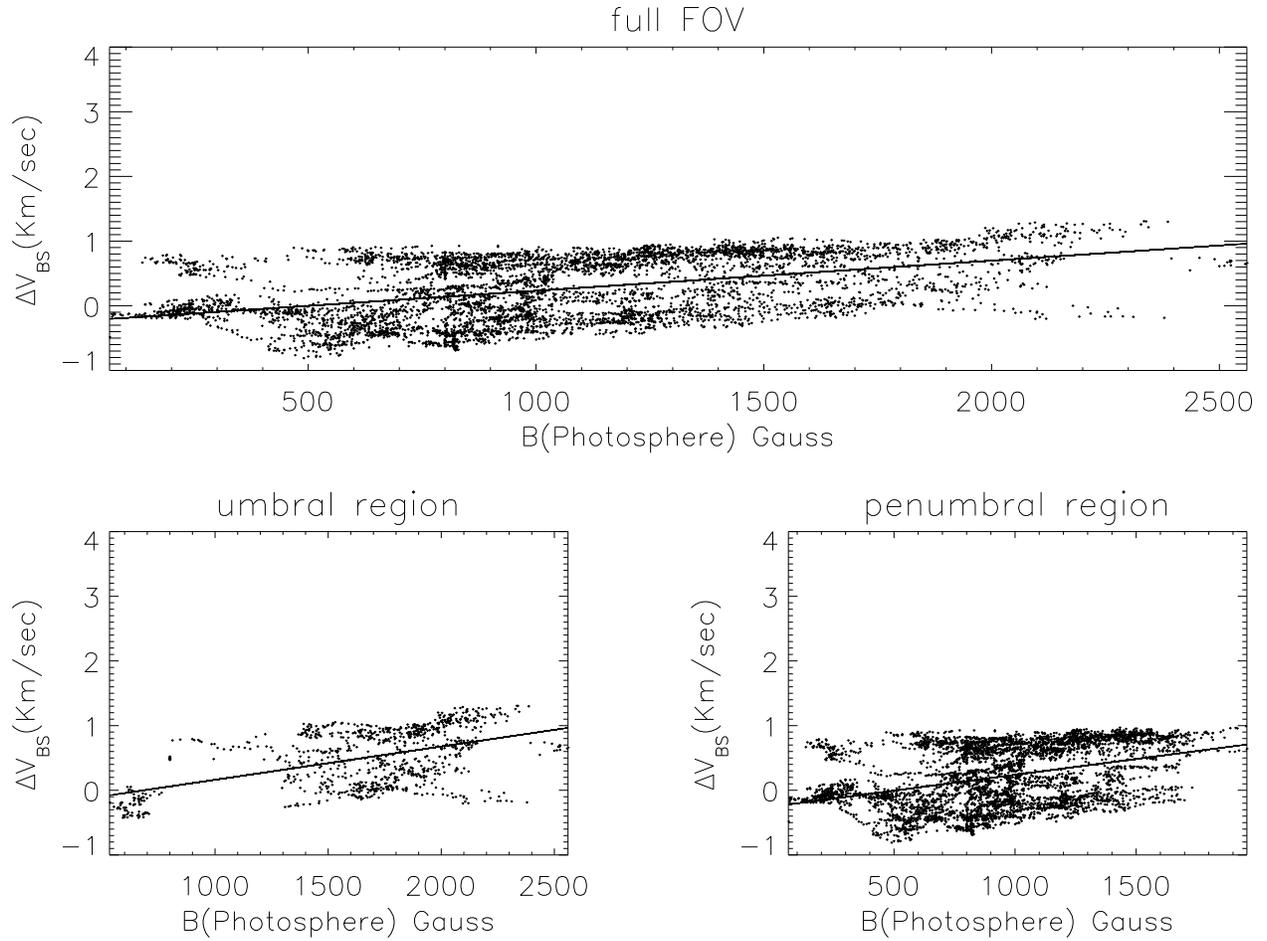}
\caption{ Same as Fig. \ref{fig:vgradhalpha_full} but 
with the limited spectral range about the line core considered for
calculating velocity gradients to avoid the influence of blending due to Co I
line.
The axes scales are kept same as Fig. \ref{fig:vgradhalpha_full} to 
indicate that the velocity gradients are smaller compared to the case when the 
full Stokes $I$ profile is considered.}
\label{fig:vgradhalpha_part}
\end{figure}

\begin{figure}
\includegraphics{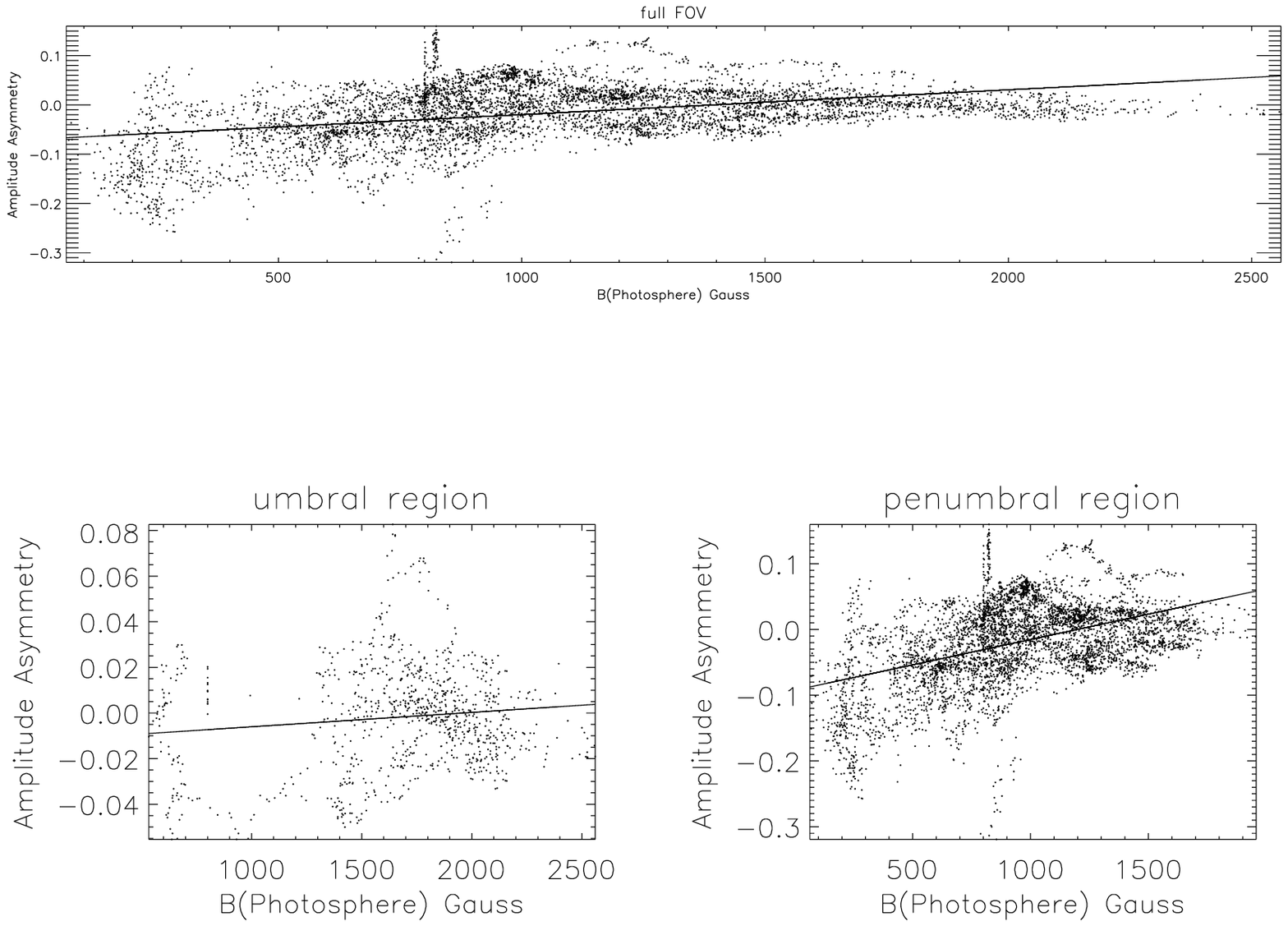}
\caption{Plots of amplitude asymmetry of Stokes V profiles of Fe I line v/s the 
total field strength (photospheric). The solid lines are the linear fits
to the data points.}
\label{fig:ampasym_fe}
\end{figure}

\begin{figure} 
\includegraphics{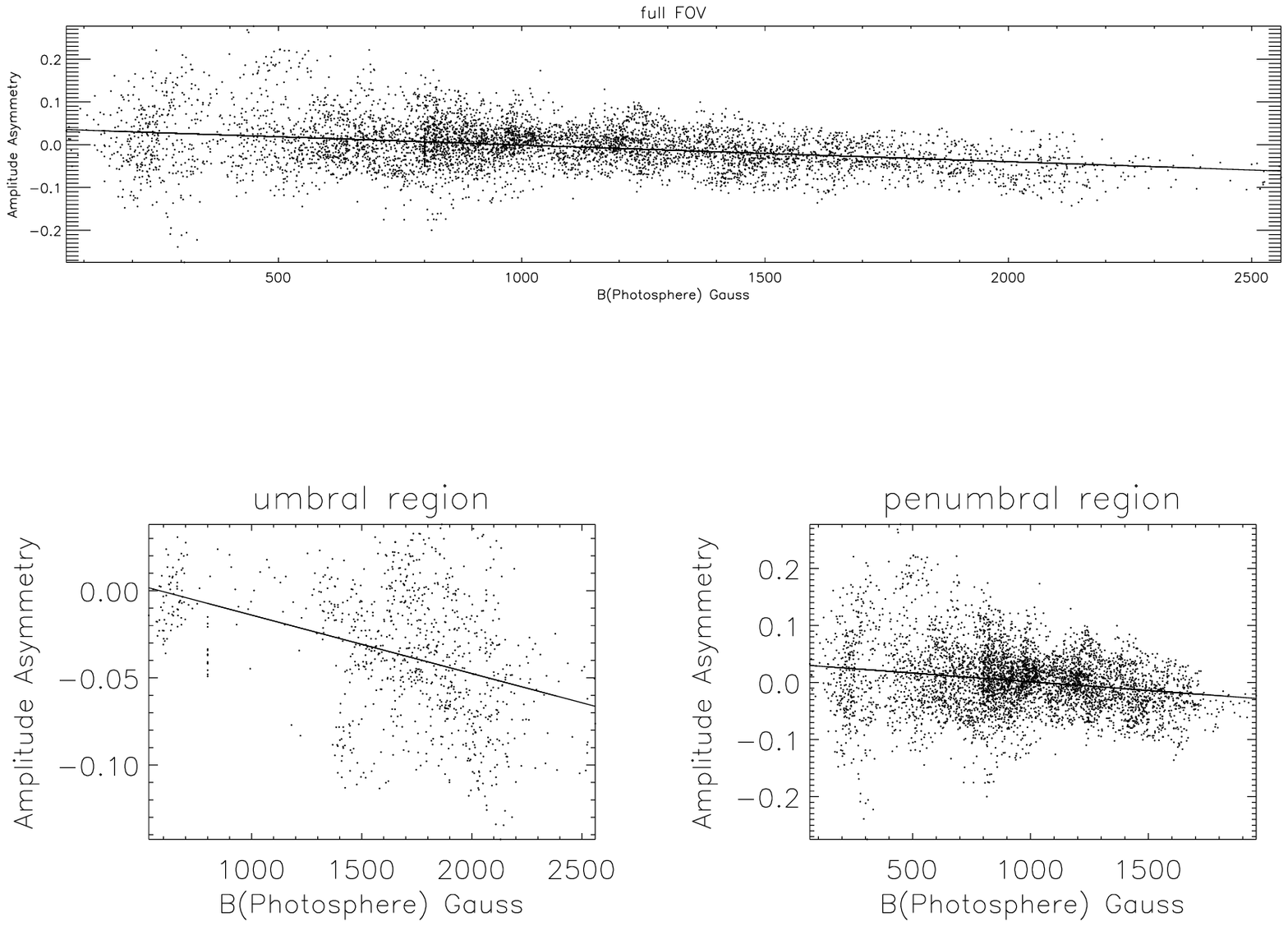}
\caption{Plots of amplitude asymmetry of Stokes V profiles of $H_{\alpha}$ line
v/s the total field strength (photospheric). The solid lines are the linear fits
to the data points.}
\label{fig:ampasym_hi}
\end{figure}

As discussed in section \ref{sect:losmag}, the LOS magnetic field measured in 
chromosphere is in general weak compared to its photospheric counterpart. The
weakening of the chromospheric field is much faster for the corresponding strong
photospheric field. The magnetic field strengths observed in the umbral 
chromosphere are much weaker than those expected from the extrapolation of
the photospheric magnetic field.  For instance, the field strength inferred 
through  $H_{\alpha}$ observations is about 400 G where as the field strength 
obtained through the extrapolation of the observed photosperic field under 
potential field approximation to an height of 2000 km is about 1000 G 
(assuming that the height of formation of $H_{\alpha}$ is about 2000 km). 
Various possibilities have been discussed which can cause the weaker fields 
observed in the umbral chromosphere. 
The most probable ones are the fast divergence of the stronger fields when 
they propagate upward in the atmosphere 
\citep{bala04} and  ambiguity in the height resolution of  
$H_{\alpha}$ magnetic sensitivity  which may be
photospheric and/or chromospheric depending on the region of observation
\citep{hector04}. If former is the reason, then it can explain the observed
properties of both velocity and the magnetic fields presented in this paper. 
This is because, as shown in Fig. \ref{fig3}, observed chromospheric fields
are systematically weaker at the locations where strong photospheric fields
are observed. If the weaker field strengths observed at the umbral chromosphere
are truly of solar origin then this implies that the umbral fields
decrease more rapidly with height compared to penumbral fields.
Rapid decrease in field strengths cause rapid increase in velocity with
height as there is a small upflow in the umbral photosphere. 
Observations also show that the velocity increases more rapidly in the
umbral region compared to penumbral region (section \ref{sect:velgrad}).
Earlier observations by \citet{gosain03} have also indicated the weakening
of the magnetic field which is larger for stronger fields.
In their observations quicker weakening of the stronger fields is not apparent,
probably, because of the lines (Mg I  b1 and b2 at $\lambda5173$ and
$\lambda5184$) used to infer
chromospheric magnetic field that originate at the lower chromosphere.
While the observations based on $H_{\alpha}$ presented in this paper
as well as earlier by \citet{bala04,hanaoka05} show clearly that the
strong fields weaken quickly with height because in these works, to infer
chromospheric magnetic field only the spectral region close to its line
core is considered which samples mostly the higher layers of the 
chromosphere \citep{rutten07}.
Hence there is a possibility that the weaker LOS field strengths observed in 
the umbral chromosphere are caused due to the faster divergence of the stronger 
fields. However, we would like to caution that the reliability of the COG
method in estimating the LOS magnetic field strengths discussed in this paper 
(section \ref{sect:losb}) is for simple solar atmospheric model. 
But, in reality the solar atmosphere may be complicated.
More studies on the magnetic and velocity response functions for $H_{\alpha}$ 
in different regions will help in better interpretation of the observations. 
Simultaneous multiline spectropolarimetry, which includes $H_{\alpha}$ and at 
least one more line which is formed at the chromosphere (preferably Infrared 
$Ca$ Triplet line at $\lambda$8542) and a photospheric line, is needed to get 
further insight into the physical processes that take place in the chromosphere.

\acknowledgments
Hector Socas-Navarro and Han Uitenbroek are thankfully acknowledged for
providing us MELANIE and RH codes. We thank K. S. Balasubramaniam for 
his critical comments on our earlier version of this paper. We also thank
S. P. Rajaguru for useful discussions regarding the connection between
velocity and magnetic field gradients and B. Ravindra for providing us the
extrapolated magnetic field strengths under potential field approximation.
Thanks to the help of Devendran and Hariharan during the observations.
The sunspot image was obtained from SOHO/MDI data acrhive. SOHO is a mission 
of international cooperation between ESA and NASA.

\clearpage

\clearpage

\clearpage

\clearpage

\clearpage

\clearpage

\clearpage

\clearpage

\clearpage

\end{document}